\documentclass[dvips]{arxstspdf}

\volume{25}
\issue{1}
\pubyear{2010}
\firstpage{41}
\lastpage{50}
\doi{10.1214/10-STS320}



\begin{document}
\begin{frontmatter}

\title{Make Research Data Public?---Not Always so Simple: \textit{A Dialogue for
Statisticians and Science Editors}}
\runtitle{Make Research Data Public?}

\begin{aug}
\author[a]{\fnms{Nell} \snm{Sedransk}\ead[label=e1]{sedransk@niss.org}},
\author[b]{\fnms{Linda J.} \snm{Young}\corref{}\ead[label=e2]{ljyoung@ufl.edu}},
\author[c]{\fnms{Katrina L.} \snm{Kelner}\ead[label=e3]{kkelner@aaas.org}},
\author[d]{\fnms{Robert A.} \snm{Moffitt}\ead[label=e4]{moffitt@jhu.edu}},
\author[e]{\fnms{Ani} \snm{Thakar}\ead[label=e5]{thakar@skysrv.pha.jhu.edu}},
\author[f]{\fnms{Jordan}~\snm{Raddick}\ead[label=e6]{raddick@pha.jhu.edu}},
\author[g]{\fnms{Edward J.} \snm{Ungvarsky}\ead[label=e7]{eungvarsky@gmail.com}},
\author[h]{\fnms{Richard W.} \snm{Carlson}\ead[label=e8,text=rcarlson@ciw.\\ edu]{rcarlson@ciw.edu}},
\author[i]{\fnms{Rolf} \snm{Apweiler}\ead[label=e9]{apweiler@ebi.ac.uk}},
\author[j]{\fnms{Lawrence~H.}~\snm{Cox}\ead[label=e10]{lgc9@cdc.gov}},
\author[k]{\fnms{Deborah} \snm{Nolan}\ead[label=e11]{nolan@stat.berkeley.edu}},
\author[l]{\fnms{Keith} \snm{Soper}\ead[label=e12,text=keith\_soper@\\ merck.com]{keith\_soper@merck.com}}
\and
\author[m]{\fnms{Cliff} \snm{Spiegelman}\ead[label=e13]{cliff@stat.tamu.edu}}
\runauthor{N. Sedransk et al.}

\affiliation{National Institute of Statistical Sciences (NISS),
University of Florida, American Association for the Advancement of Science,
Johns Hopkins University,
Johns Hopkins University,
Johns Hopkins University,
Northern Virginia Capital Defender Office,
Carnegie Institution of Washington,
European Bioinformatics Institute of the European Molecular Biology Laboratory,
National Center for Health Statistics, University of California at Berkeley,
Merck, Inc.
and Texas A\&M University}

\address[a]{Nell Sedransk is Associate Director and Professor of Statistics, National
Institute of Statistical Sciences (NISS), 19 T. W. Alexander Drive, Research Triangle Park, North Carolina 27709, USA \printead{e1}.}
\address[b]{Linda J. Young is Professor of Statistics, Department of Statistics, University of Florida, 408 McCarty Hall C, P.O. Box 110339,
Gainesville, Florida 32611-0339, USA \printead{e2}.}
\address[c]{Katrina L. Kelner is Editor in Chief American Association for the Advancement of Science and Editor,
The Journal Science Translational Medicine,
American Association for the Advancement of Science, 1200 New York Avenue NW, Washington, DC 20005, USA \printead{e3}.}
\address[d]{Robert A. Moffitt is Krieger--Eisenhower Professor of Economics, Department of Economics, Johns Hopkins University, 449 Mergenthaler Hall, 3400 N. Charles Street, Baltimore, Meryland 21218, USA \printead{e4}.}
\address[e]{Ani Thakar is Research Scientist, Center for Astrophysical Sciences, Johns Hopkins University, 3701 San Martin Dr., Baltimore, Maryland 21218-2695, USA \printead{e5}.}
\address[f]{Jordan Raddick is Education and Public Outreach Specialist, Department of Physics and Astronomy,
Johns Hopkins University, Bloomberg 388, 3400 N. Charles Street, Baltimore, Maryland 21218-2686, USA \printead{e6}.}
\address[g]{Edward J. Ungvarsky is Capital Defender, Northern Virginia Capital Defender Office, P.O. Box 17010, Arlington, Virginia 22216, USA \printead{e7}.}
\address[h]{Richard W. Carlson is Staff Member, Department of Terrestrial Magnetism, Carnegie Institution of Washington, 5241 Broad
Branch Road NW, Washington, DC 20015-1305, USA \printead{e8}.}
\address[i]{Rolf Apweiler is Joint Team Leader, Panda Group, Protein and Nucleotide Database Group, European Bioinformatics Institute
of the European Molecular Biology Laboratory, Wellcome Trust Genome Campus, Hinxton, Cambridge, CB10 1SD, United Kingdom \printead{e9}.}
\address[j]{Lawrence H. Cox is Associate Director, National Center for Health Statistics, 3311 Toledo Rd, Hyattsville, Maryland 20782, USA \printead{e10}.}
\address[k]{Deborah Nolan is Professor of Statistics, Department of Statistics, University of California at
Berkeley, 367 Evans Hall 3860, Berkeley, California 94720-3860, USA \printead{e11}.}
\address[l]{Keith Soper is Senior Director, Biometrics Research, Merck, Inc., WP 37C-305, 770 Sumneytown Pike, West Point, Pennsylvania 19486-0004, USA \printead{e12}.}
\address[m]{Cliff Spiegelman is Professor of Statistics and Toxicology, Department of Statistics, Texas A\&M University, 3143 TAMU,
Blocker Building, Room 447, College Station, Texas 77843-3143, USA \printead{e13}.}
\end{aug}\vspace*{-3pt}

\begin{abstract}
Putting data into the public domain is not the same thing as making
those data accessible for intelligent analysis. A distinguished group of
editors and experts who were already engaged in one way or another with
the issues inherent in making research data public came together with
statisticians to initiate a dialogue about policies and practicalities
of requiring published research to be accompanied by publication of the
research data. This dialogue carried beyond the broad issues of the
advisability, the intellectual integrity, the scientific exigencies to
the relevance of these issues to statistics as a discipline and the
relevance of statistics, from inference to modeling to data exploration,
to science and social science policies on these issues.\vspace*{-3pt}
\end{abstract}

\begin{keyword}
\kwd{Data availability}
\kwd{data reuse}
\kwd{forensics statistics}
\kwd{data
policy}
\kwd{journal policies on data}
\kwd{proteomics statistics}
\kwd{geochemical
data base}
\kwd{sky survey}
\kwd{astronomy data}.\vspace*{-12pt}
\end{keyword}

\end{frontmatter}

\section*{Motivation}

For the highest scientific standards to be upheld, many of us agree that
scientific findings and pronouncements need to be supported with facts.
This is especially so for findings that directly impact the health,
well-being or freedom of people. For science to be open requires
provision of reasonable access to data and metadata, together with clear
statements of relevant assumptions, experiments, and inferences. Yet
data today are often not available or are provided with inadequate
context. Making data from scientific articles widely accessible requires
grappling with many problems, from ethical to technological to fiscal.
The workshop that prompted this paper attempted to summarize common
threads into principles for proceeding toward openness in science. As
scientific journals and professional societies grapple with these
issues, attention to relevant statistical issues must be kept in focus.
At the same time, the rich opportunities to participate in and
contribute to scientific data-sharing pose new challenges to the
statistics profession---challenges that are simultaneously being taken
up by other computational disciplines. What are these challenges and
opportunities?

\textit{First} is the challenge \textit{to act within the profession} to
establish criteria that define one or more levels of data availability
for data published by professional statistical journals.

\textit{Second} is the challenge \textit{to support scientific communities} by
defining statistical criteria that expand the scope of usefulness of
data made available by those communities and their journals.

\textit{Third} is the challenge \textit{to provide sound statistical algorithms,
modules and software} directly to the large, cooperatively generated
databases being established and interlinked by scientific communities
that depend upon sharing rare or costly data.

\textit{Fourth} is the challenge \textit{to identify and/or to create special
tools} specific to the new multifaceted research possibilities, and
simultaneously \textit{to identify} (new) \textit{scientific pitfalls}
arising from multiple use of rare or compiled data.

\textit{Fifth} is the challenge \textit{to advance statistical education and use
of appropriate statistical methodology} for important data in new venues
and for new audiences. This is in addition to the traditional statistics
courses and science courses at all levels from pre-college through
graduate school.

This article began with a workshop where statisticians engaged with
science and social science researchers and editors who are grappling
with the consequences of making scientific research data publicly
available. The workshop, held February 8--9, 2008, at the offices of the
AAAS in Washington, DC, was the first in what is anticipated to be a
series developed under the guidance of the National Institute of
Statistical Sciences's (NISS's) National Priorities Committee. The
format chosen for this workshop was an exploration in dialogue between
statistics and the sciences. This paper reports issues raised and
discussed in that workshop, which were deliberately circumscribed to
focus on data rather than data processing or archiving, leaving these
important issues to another forum.

\section*{Dialogue with Science Editors and~Experts}

The science journal editors opened the dialogue with experiences from
their disciplines' perspectives on the benefits and challenges in making
research data available. Advantages and difficulties varied\break among the
disciplines, each having different implications for Statistics.
Summaries of several of these are presented here.

\subsection*{Katrina Kelner, Deputy Editor for Life~Sciences,~\textit{Science}}

\textit{Science} adopted the following data access policy: ``After
publication, all data necessary to understand, assess, and extend the
conclusions of the manuscript must be available to any reader of
\textit{Science}.'' This policy resulted from discussion of the
benefits, the risks and the practicalities of making data publicly
available. When publishing a scholarly paper, journals take on the
responsibility of hosting and disseminating the data underlying the
conclusions\textit{. Science} has had an evolving policy on this score.
Data can be housed in the main text of the paper itself, in the
supplementary online material, in public repositories, or in rare cases
on the author's web site. Ambiguities remain in implementation and
enforcement of these policies. For example, one question to decide is:
What are ``data''? Does the term data include raw data (X-ray film),
counts (from radioactivity-measurement instruments), processed data
(sequence data, classified images, behaviors recorded and counted on a
video), summaries of all data values, or tables and figures prepared for
publication? Defining ``data'' is insufficient without an equally
careful definition of metadata, that is, the scientific context within
which the data can be understood and interpreted.

Obstacles to data sharing may be technical, practical and/or legal.
Technical issues include discipline-specific conventions, computer code
and machine-specific embedded preprocessing software. Practical
obstacles include nonuniformity of format, massive size of data sets and
lack of public databases. Legal obstacles include privacy laws,
multi-author ownership of the data, material transfer agreements and
proprietary data sets. As important as any of these is an author's
desire to capture (first) rewards from his/her data. (See the recent
National Academy of Science report by COSEPUP for thorough discussion of
these issues.)

\begin{proposition*} The traditional print-based scientific paper is no
longer the optimal format for presenting peer-reviewed scientific
results.
\end{proposition*}

\subsection*{Robert Moffitt, Editor of \textit{The American Economic~Review}}

Data sharing has a fairly recent history in economics, and the field is
moving gradually toward more extensive sharing. Economics has some
unique features that distinguish it from other disciplines. For example,
the majority of data used by economists is publicly available because
they are provided to all researchers by the government. Nevertheless, a
small but growing number of economists use confidential information from
firms or other organizations which have legitimate rights to privacy of
information but which often will allow selected researchers to analyze
their data. Studies of this kind constitute about 10 percent of
submissions to leading economic journals; and data sharing requires the
permission of the owning organization, which is not always forthcoming.
However, the main issue in economics for the 90 percent of research that
uses publicly available data is not data sharing per se but
rather replication of results. Replication requires knowing how the data
were manipulated in the process of conducting the analysis. Data may be
subsetted, imputed, trimmed; outliers may be removed; variables and
classifications may be created in the process of conducting the
research. The actual analysis files may comprise a sample, even a random
sample, of a much larger database; in consequence, the keys to
replication require not only the sampling algorithm but the specific
sample. For reasons that are not always obvious, different researchers
often obtain different results when analyzing data from the same master
file; and it is sometimes the case that researchers' results are
sensitive to the choice of methods used. Other studies may be in actual
error.

This focus on replicability and robustness testing has indirectly led to
this demand for data-sharing, a relatively recent phenomenon in
economics. The professional peer-reviewed journals in economics have
taken a leadership role in its promotion. \textit{The American Economic
Review} now requires authors of accepted papers to provide both their
data sets and the programs used to manipulate the data for posting on
its web site. This policy has been well received by the profession, and
compliance has been 100\%, although this is as a percent of articles
based on nonproprietary data. Most of the other leading journals in
economics have recently followed the lead of \textit{The American
Economic Review} and are now similarly requiring that authors of
accepted papers provid their data and programs. Progress is slow but
steady and the desired ``Culture of Replication'' and data sharing that
would be optimal is still quite a way off.

\subsection*{Ani Thakar and Jordan Raddick, Johns Hopkins University,
\textit{Sloan Digital Sky Survey} (SDSS)}

The whole goal of the SDSS, funded by the National Science Foundation,
is to share data with the World. The ``World'' means elementary school
students and teachers, college students and teachers, foremost
researchers, foremost astronomers and the general public---that is,
anyone seriously interested or casually intrigued by the objects in the
sky. For years, astronomers took photos, and astronomers looked at them.
Now instruments are digital, yielding petabytes of data (the equivalent
of tens of thousands of CDs). The number of visible galaxies has gone
from 200,000 to 200,000,000. After two years of availability to
astronomers for their research, these data become public. At today's
rate, the data doubles every two years so that 50\% is available to the
public at any time. Techniques (statistical and otherwise) built for the
``old days'' cannot drink from this fire hose of data. The only feasible
way to access this data is (or could be) online, making the internet
potentially the world's best telescope. Even so, it has data on every
part of the sky, in every measured spectral band (optical, X-ray,
radio). It sees as deep as the world's best instruments; better still,
it is optimal when you are awake, and the observing conditions are
always great.

Sharing data with the public is also an access to learning, whether at
age 6 or 60, whether scientist or nonscientist, whether amateur or
superb researcher. Everyone has free choice to look at any star or any
galaxy, to ask questions and to form opinions. Users appreciate the
trust implied by giving them real data; they also need public tools that
work at their level. There is a huge audience for ``Citizen Science'';
and this is a rare chance to show science in action. However, the
research problem must be authentic, and the tools must yield bona fide
interpretations.

\subsection*{Edward Ungvarsky, Capital Defender, Northern Virginia Capital
Defender Office}

While forensic ``evidence'' is hardly a novel concept in criminal law,
forensic ``science'' is relatively new to criminal cases. Previously,
police technicians, not the scientifically or mathematically
trained,\break
opined about the significance of forensic evidence.\footnote{Simon Cole, \textit{Suspect Identities: A History of Fingerprinting and
Criminal Identification} (Harvard Univ. Press 2001); Arthur Conan
Doyle, \textit{The Bascombe Valley Mystery. The Adventures of
Sherlock Holmes} (1891).} Because their
opinions were accepted at face value, these technicians long controlled
the presentation of forensic evidence in court.\footnote{Simon Cole, \textit{Suspect Identities: A History of Fingerprinting and
Criminal Identification} (Harvard Univ. Press 2001).}

In the late 1980s, science moved into criminal\break courtrooms with the
advent of forensic DNA evidence. There was great scientific debate about
the utility of the evidence,\footnote{Richard C. Lewontin and Dan Hartl, Population genetics in
forensic DNA typing. \textit{Science} \textbf{254} 1745 (1991); Ranajit Chakraborty
and
Kenneth K. Kidd, The utility of DNA typing in forensic work. \textit{Science}
\textbf{254}  1735 (1991).} lengthy court admissibility
hearings\footnote{\textit{People v. Castro}, 545 N.Y.S.2d 985 (N.Y. Sup. Ct.
1989); \textit{United States v. Yee}, 134 F.R.D. 161 (N.D. Ohio
1991).} and national scholarly attention paid to the benefits and
limits of the evidence.\footnote{National Research Council, DNA Technology in Forensic Science
(1992); National Research Council, The Evaluation of Forensic DNA
Evidence (1996).} While testing modalities have changed
since it was first used in a courtroom in 1986, typical\footnote{``Typical'' DNA analysis excepts more novel, unproven applications
to mitochondrial, Y-STR and low copy number DNA, as well as familial
searching. Frederika A. Kaestle, Ricky A. Kittles, Andrea L. Roth and
Edward J. Ungvarsky, Database limitations on the evidentiary
value of forensic mitochondrial DNA evidence. \textit{Am. Crim. L. Rev.} \textbf{43} 53
(2006); \textit{Yeboah v. State}, No. A07-0739 (Minn. Ct. App. May
13, 2008); \textit{People v. Espino}, No. NA076620 (Super. Ct. L.A.
Co. Cal. March 18, 2009); Bruce Budowle, Arthur J. Eisenberg and Angel van
Daal, Validity of low copy number typing and applications to
forensic science. \textit{Croat. Med. J.} \textbf{50} 207 (2009); Frederick R. Bieber,
Charles H. Brenner and David Lazer, Finding criminal through
DNA of their relatives. \textit{Science} \textbf{312} 1315 (2006); Jules Epstein,
``Genetic Surveillance''---The bogeyman response to familial
DNA investigations. \textit{U. Ill. J. L. Tech. \& Policy} (2009); David R.
Paoletti, Travis E. Doo, Michael L. Raymer and Dan E. Krane,
Assessing the implications for close relatives in the event
of similar but nonmatching DNA profiles. \textit{Jurimetrics} \textbf{46} 161 (2006).} forensic
DNA analysis is considered by many to be the ``gold standard'' of
scientific evidence.\footnote{William C. Thompson, Tarnish on the  ``Gold Standard'':
Recent problems in forensic DNA testing. \textit{Champion} \textbf{Jan./Feb.} 10 (2006);
Michael J. Saks and Jonathan J. Koehler, The coming paradigm
shift in forensic identification science. \textit{Science} \textbf{309} 892 (2005);
National Research Council, Strengthening Forensic Science in the United
States: A Path Forward (2009).} Forensic DNA is customarily admitted in court
with an associated product rule-based ``random match probability
estimate,'' or the likelihood that a randomly chosen, unrelated person
in a particular population would have a particular DNA profile that has
also been found in an evidence sample.\footnote{\textit{United States v. Porter}, 618 A.2d 629, 640 (D.C. 1992); \textit{United
States v. Cuff}, 37 F. Supp. 2d 279, 282 (S.D.N.Y. 1999); David H. Kaye
and George F. Sensabaugh Jr., Reference guide on DNA evidence. In
\textit{Reference Manual on Scientific Evidence}, 2nd ed. (Federal Judicial Center
2000) 545; National Research Council, DNA Technology in Forensic Science
(1992); National Research Council, The Evaluation of Forensic DNA
Evidence (1996).}

Forensic DNA analysis has advanced to the point where
random-match-probability estimates in the 1-in-quintillions are
routinely reported,\footnote{\textit{People v. Nelson}, 48 Cal. Rptr.3d 399 (Cal. Ct. App.
2006), aff'd, 185 P.3d 49 (Cal. 2008).} particularly in cases where a suspect is
identified by trawling a federal database of over seven million arrestee
and convicted offender profiles.\footnote{At over seven million profiles, this database is large enough that
one would expect to see a profile with a random-match-probability
estimate of 1-in-24 trillion (over 35 times the Earth's population)
appear twice. \url{http://www.fbi.gov/hq/lab/codis/clickmap.htm}.}  Such figures call for a
re-appraisal of the assumptions underlying random-match-probability
estimates via statistical study of the data contained in this and
similar state\break databases.\footnote{Edward J. Ungvarsky, What does one in a trillion
mean? \textit{Genewatch} \textbf{Feb.} (2007); Erin Murphy,  The new
forensics: Criminal justice, false certainty, and the second generation
of scientific evidence. \textit{Cal. L. Rev.} \textbf{95} 721 (2007); Keith Devlin,
 Damned lies. \textit{MAA Online} \textbf{Oct.} (2006); David H. Kaye,
Trawling DNA databases for partial matches: What is the FBI
afraid of? \textit{Cornell J. L. \& Public Policy} \textbf{19} 1 (2009); Laurence D.
Mueller,
Can simple population genetic models reconcile
partial match frequencies observed in large forensic databases? \textit{J.
Genetics (India)} \textbf{87}  101 (2008); Y.~S. Song, Ananda Patil, Montgomery
Slatkin  and Erin Murphy,  The Average probability that a cold
hit in a DNA database results in erroneous conviction. \textit{J. Forensic Sci.}
(2009).}  Law enforcement's strenuous resistance
to any effort to access their databases---whether by scholars
interested in scientific study of the wealth of data, or by lawyers
seeking to identify the perpetrator of the offense as the source of an
unknown crime scene profile---should cease.\footnote{David H. Kaye,  Trawling DNA databases for partial
matches: What is the FBI afraid of? \textit{Cornell J. L. \& Public Policy} \textbf{19}
1 (2009);
Edward J. Ungvarsky, What does one in a trillion
mean?  \textit{Genewatch} \textbf{Feb.} (2007); Bruce Budowle et al., Partial
matches in heterogeneous offender databases do not call into question
the validity of random match probability calculations. \textit{Int'l J.
Legal Med.} \textbf{123}  59 (2009); Martha Neil,  FBI ordered to do DNA
search to help suspect in rape--murder case, available at
\url{http://www.abajournal.com/index.php?/news} (Feb. 3, 2009).}

Federal law enforcement also maintains forensic databases for many other
types of physical evidence that are routinely used in criminal
prosecutions and are admitted with statements of source attribution
without recognition of probabilistic limits.\footnote{National Research Council, Strengthening Forensic Science in the
United States: A Path Forward (2009).} These other
databases too are withheld from the type of scholarly investigation
undertaken to ensure the accuracy, reliability and validity in
scientific disciplines. After some high-profile forensic
misidentifications and an in-depth scientific review, new emphasis has
been placed on the need for research to address the fundamental
scientific validity of these identification disciplines.\footnote{National Research Council, Strengthening Forensic Science in the
United States: A Path Forward (2009).} Rather
than technicians simply opining, without any statistical basis, that
forensic evidence matches a particular source to the exclusion of all
others, research demonstrating the probabilistic likelihood of such
matches is now recommended. If done correctly, this research should
convert forensic ``evidence'' into forensic ``science.''\footnote{National Research Council, Strengthening Forensic Science in the
United States: A Path Forward (2009).}
Recommendations for access to criminal databases and for review of
forensic evidence bode well for a new age of scientific engagement with,
and improvement of, the criminal justice system.

\subsection*{Richard W. Carlson, Editor of \textit{Earth \& Planetary~Letters}}

An observer from outside the earth sciences might view earth science as
a single discipline. A geologist or geochemist or geophysicist
recognizes the interdisciplinary distinctions. Data handling in these
sub-disciplines has evolved along quite different paths reflecting the
types of data involved and the data analysis needs of the different
disciplines. In seismology, the basic data, seismograms, are quite
simple, with minimal associated metadata, and hence are relatively
easily archived. Seismic imaging of Earth's interior, however, relies on
the analysis of tens of thousands of seismograms recorded from widely
spaced localities on Earth's surface. The need for analysis of large
data sets contributed to the seismology community organizing the
Incorporated Research Institutions for Seismology (IRIS) whose Data
Management Center now archives, and freely serves, a large fraction of
the world's seismic data.

In geochemistry, the basic data are simple (e.g., elemental composition
of rock samples), but the metadata are complex and include such
information as sample name, collection locality, rock type, analytical
techniques used, and the precisions and detection limits of the
techniques. Until the early 1970's, individual rock analyses were hard
won. The small quantities of data and metadata could be published in
paper form in the journals of the discipline. With the advent of
automated geochemical instrumentation, data quantities rapidly expanded
beyond what journals were willing to publish, leading to the publication
of data ``summaries,'' with the raw data being retained by the author,
often inaccessible to other researchers. Electronic online supplements
removed this barrier to the publication of large, complex, digital
tables, but the way that such data tables are served by the journals of
the discipline is little improved over the paper-publishing era.

Attempts to place the published data in relational databases were
initiated only in the last decade. One example is the EarthChem
(\href{http://www.earthchem.org}{www.earthchem.org})\break database
that freely serves geochemical data for\break
nearly 600,000 rock samples. EarthChem accommodates all essential
metadata for each rock sample and is applicable for both individual rock
data and for the mineral constituents of many rock samples. Dynamic,
interactive, web-based user interfaces access these data to supply
integrated information\break about an individual rock sample with references.
This has opened new research opportunities for\break cross-disciplinary
analyses of well-studied,\break well-characterized, individual rocks. With the
addition of fairly simple statistical summaries and calculations, this
is also changing the education and next generation of scientists. With
standards for metadata, unique sample identification, map interfaces,
visualization tools for data selection (rock sampling) and integrated
tools for data analysis, both students and researchers can explore this
multidisciplinary world. The attraction is easy to explain. The rapid
growth of a community of regular users is proof.

\subsection*{Rolf Apweiler, Past President, Human Proteome Organisation
(HUPO)}

Proteomics is an expensive technology, based on mass spectrometer
equipment, and dependent on software to create interpretable data from
the raw instrument output. There are at least five reasons for making
proteomics data available: (1) Science has been built upon the knowledge
and sharing of information. (2)~Data users are not necessarily the best
analysts nor the best developers of analytic tools. (3) Meta analysis of
data can recycle previous data and findings for new tasks. (4) Sharing
data allows independent review of the findings. (5) Simple economics.
``Information, no matter how expensive to create, can be replicated and
shared with little or no cost'' (Jefferson). Simply sharing data is not
enough.

Available data is not necessarily accessible data. When data are only
made available as arbitrarily formatted tables, they carry important
limitations. Without source data, true peer review and validation is not
possible; with very little raw material, testing and retesting may be
impossible. The result of the first may be large numbers of publication
without validation. The result of the second may be data hoarding to
protect a scientist's own line of research. A second limitation arises
from the automated preprocessing, differences in embedded software of
different manufacturers' instruments, making objective technique
comparisons difficult or unachievable. Accessibility requires
infrastructure, community supported standardization, controlled
vocabularies and ontologies, minimum reporting requirements, and
publicly available online repositories.\break Bioinformatics grew up alongside
the internet, and this is reflected in the successful online data
sharing mechanisms already in place in the life sciences.

The final goal is cross-domain integration and validation. How do we
make this all happen? First are journal guidelines that heavily
influence the decisions made by authors. By first requesting and then
mandating data submission to established repositories, journals provide
an important ``stick.'' Gaining editorial board and community consent is
not a foregone conclusion. Second, funders' support and guidelines
contribute both ``sticks and carrots.'' Third, the data repositories
must be freely available and reliable. Feedback loops need to be
established to ensure the accumulated data flows back to the user
community. While there are successes, there are also authors who will
choose whenever possible to submit their papers without the burden of
providing truly accessible data.

At the conclusion of the editors' and experts' presentations, one posed
two questions:\vspace*{4pt}

\begin{center}
\textit{``But what does Statistics have to do with this?''}

\textit{``And, what does any of this have to do with Statistics?''}
\end{center}

\section*{In Reply}

This article is \textit{not} intended to provide the answers,
\textit{nor even} to identify all the opportunities, although some are
referenced. Rather its intent is to stimulate responses from the various
members and organizations within the statistical community to respond to
the challenges that this complex issue poses for science. At stake are
the ideals of openness and preservation of scientific integrity. At risk
is the representation of faulty reasoning as science, especially where
deep technical skill is required to discern the critical, logical or
technical flaw. One role of statistics is to clarify the reasoning and
to support the scientific interpretation by meeting the challenges posed
by the science editors and researchers at the workshop.

\section*{Challenge \#1---To Act within the~Profession }

\subsection*{Statistics Journals}

Journals in other sciences have struggled and continue to struggle with
policies on making publicly available the scientific data on which the
articles they publish are based. The key issues apply equally for
statistics journals.

Which data are to be made available: Original data (with
de-identification of individual subjects)? Aggregate data? At what scale
of aggregation? Preprocessed data? After how much pre-processing?\break Mixed
original and synthetic data? Subsamples from original [massive] data?

Key issues also include the mechanics of availability: Who will maintain
the data? Where? In what form? With what metadata? For how long? At what
cost? Paid by whom?

They also include preservation of scientific integrity of data, security
and privacy: How can data be protected from alteration, deletion or
other distortion? What about mischievous or even malevolent reanalysis?
What are the IP rights of data providers when their data is reused? What
about citation permission, caveats, credit? Who has the responsibility
to make data available from interdisciplinary research?

Privacy issues take on a new life when data access expands from the
primary researchers (data gatherers) to the public. Once in a more
public domain, abundant auxiliary sources of information might be joined
with the original [de-identified] data to decipher individual identities
of human subjects or of proprietary information. What guarantees of
privacy can be given to study subjects or to individual suppliers of
data, especially proprietary or confidential data?

Journals themselves have additional issues: How do review
responsibilities change when available data is reused by a new author?
Do reviewers need access to original data prior to acceptance and will
reviewers be willing to examine that data? What about papers presenting
conflicting inferences from a single data set? Who will pay for the data
archives? Will authors publish their best work in journals that require
data availability (and the additional work by the authors to prepare the
data for archiving)? For example, does an archeologist have to provide
data obtained from years of investigation with the first paper published
using the data?

The struggles to arrive at answers to these questions become worthwhile
when the data become truly accessible to additional users to answer
scientific questions that could or would not be addressed otherwise and
also become available to statisticians and others to develop and
validate new statistical methodology.

Is data availability impossible? Probably, no. For \textit{The Annals of
Applied Statistics} (\textit{AOAS}), authors are strongly encouraged to
make data used in their papers available. Data sets as well as software
and extensive mathematical derivations are reviewed with the paper. When
a paper is accepted for publication, these supplementary materials are
placed in the dedicated \textit{AOAS} Data and Software Archive at
StatLib. Numerous statistics journals, such as \textit{Biometrics} and
the \textit{Journal of the American Statistical Association}, also
encourage authors to make their data available. The extent to which the
authors are required to make data available varies.
\textit{Biostatistics} has created the position of editor for
reproducibility, in addition to annotating all articles to indicate
availability of data and of code.

\subsection*{Responsibilities and a Caveat}

Nothing can capture everything that happened as the data were originally
being gathered or generated or even as they were originally being
analyzed. But a secondary user is going to be severely hampered or
misled if key information is missing. The federal agencies and other
organizations have long dealt with provision of sufficient information
for a responsible secondary user to know what is possible and what is
not possible to assume about the process of assembling data in the
database. For example, Tranche is a data storage and retrieval resource
for the proteomics community that allows various levels of data
annotation (Falkner and Andrews, \citeyear{Falkner2007}).

The researchers responsible for origination of the data cannot be held
accountable for the objectives or the accomplishments of secondary users
of the data. The original researchers can, however, make responsible
secondary data use possible and thereby promote further achievements.

\section*{Challenge \#2---To Support Other Sciences}

\subsection*{Statistical Criteria}

Some issues, such as completeness of scientific metadata, have
significant consequences for statistical analysis and for design of
studies incorporating publicly available data. Important specific
issues, for example, time limitations on definitions of terms and
differences in the refinement of measurements possible with different
generations of measurement equipment, can be highlighted by their impact
on the final results obtained through statistical analysis and thence on
the conclusions to be drawn. Nonetheless, since these are of primary
\textit{scientific} concern, both individual scientists and the
professional societies for their disciplines are keenly aware of these.
Identification of relevant scientific metadata must be the
responsibility of the scientists working in the field; and as an
example, in astronomy work to develop a taxonomy is well advanced.

What fewer scientists appreciate is the need for the \textit{statistical
metadata}; still fewer claim the expertise to define these. Some
scientists may find providing metadata too time consuming without a
deeper understanding of the benefits for providing it. Articulating the
questions that must be answered to ensure statistical validity of
reinterpretation or reuse of scientific data is the responsibility of
statisticians:

\begin{itemize}
\item[$\bullet$] Does it matter how the data were gathered in the first
place?
\item[$\bullet$] Does it matter that the results came from data
exploration (data mining) rather than an experimental or observational
plan?
\item[$\bullet$] Does it matter whether data points were dropped or not
or whether those same points should have been dropped?
\item[$\bullet$] Does it matter if or how missing data was imputed?
\item[$\bullet$] Does it matter how measures of variation were computed
and/or whether these can be calculated from the data?
\end{itemize}

If the first set of questions focuses on the same
experimental/analytical questions that researchers consider in their own
research, a second set of questions arise when data is borrowed,
interpolated or reused, possibly multiple times.

Will it matter if all the researchers studying a rare disease use the
identical control group once it is available through a common database?
When should data \textit{not} be replicated because it can only be
redundant and therefore is an irresponsible use of funds? Does it matter
if ``synthetic'' experimental units are created from several available
sources and combined with or treated as ``real'' observations? Does it
matter if a new data set created by sampling from several data sets
gathered for different purposes is studied instead of a new sample from
a single population? What are the implications for estimation of
variation? What are the implications for statistical high-dimensional
data analysis?

Why is any of this important to the scientific conclusions that will be
drawn?

Statisticians have the knowledge to pose the crucial questions that the
scientific researchers need to answer, and statisticians have the
experience to provide cogent and persuasive illustrations and
explanations demonstrating why these are important. Articulating these
for scientists to consider as they archive data or extract data from
archives is critical for scientists. They need to understand the extent
to which that data can be useful and valid in a new context.

Data in the public domain is not usefully available without the
capability of accessing, organizing, manipulating and [re]structuring it
for analysis and for analytic software. Ancillary (statistical) support
could also take the form of recommendations about database structures
(e.g., relational databases) that facilitate analysis or even
comparative evaluation of statistical software for those analyses.
Statisticians have taken on these roles in the past for standard
statistical methodology, especially with software performance testing
and provision of reference data sets. The challenge escalates when more
computer intensive or more complex statistical analytic tools are
considered. The separate problem of configuration of data to allow
facile transfer for sophisticated statistical analysis becomes more
important as the data sets become larger, with greater internal
complexity. Going beyond Excel spreadsheets requires both suitable data
configuration and the requisite data extraction tools. Statisticians are
well positioned to provide guidance in data structures that are amenable
to the use of sophisticated statistical methodology and to the
extraction of data for reuse in subsequent research endeavors. The
challenge to statisticians here is both to advise and to develop a
knowledge base for useful guidance.

\section*{Challenge \#3---To Provide Sound~Statistics}

\subsection*{Statistical Methodology for Aggregated Data}

Combining data, whether in large amalgamated databases or simply in
assembling from several individual investigations, presents specific
needs for development of new statistical methodology. The whole
statistical research area of meta-analysis deals with a number of these
questions (Hedges and Olkin, \citeyear{Hedges1985}; Whitehead, \citeyear{Whitehead2002});
and Bayesian
methodology often applies when previous research results are used as a
springboard to subsequent studies. Still, there are new problems for
which statistical methods have not yet been devised; three of the many
kinds of problems serve as illustrations.

One class focuses on methodology---and the consequences for statistical
analysis---for subsetting\break and/or for identifying matching cases from
multiple resources whether for the purposes of comparison or for
creation of an artificial, composite individual.

A second class focuses on synthetic experimental units themselves---that is, pseudo-experimental units that are synthesized by attaching two
different experimental units from two separate sources of [different
kinds of] information and gluing them together into a single unit with
complete information. For example, combining geochemistry on one rock
sample with physical data on a second rock of the same sort obtained by
a different researcher; Or combining consumer expenditure survey
information from one participant with savings history from a banking
study for a participant matched in terms of ethnicity, age and census
tract.

A third class focuses statistical implications of the use and repeated
reuse of particular individuals or experimental units, whether a single
ocean floor rock of a particular kind or a single family tree or a
single control group from a study of a rare medical condition. This
extends to reuse of code, of models, and of open-source models and
scientific coding.

Recognizing these needs for statistical innovation is the first
challenge, meeting these needs can follow.

\section*{Challenge \#4---To Identify New Needs and New Statistical
Tools}

\subsection*{Statistical Collaboration}

Meeting the challenge to respond to ``big science'' opportunities
requires getting involved at a fundamental level, then doing the
\textit{hard} work on the \textit{hard} problems, creating the
\textit{new theory} and \textit{new methodology} for deep, complex
science. The interconnected multidisciplinary databases offer scientists
the opportunity for investigations at a new level of complexity. This
same complexity puts new demands on the analytic process and creates new
opportunities for collaboration and the extension of high-dimensional
statistical theory and methods to new arenas.

On a still higher scale, as the scientific research goals become more
complex, more and more often they are also much broader in scope. One
prime resource that statisticians have long brought to the collaboration
table is the ability to interpret the scientific context and then to
formulate a structured approach to a complex problem successfully,
leading to sound inference from the research. All this \textit{before
data}. This statistical thinking now has a place on a much larger scale
to provide a statistically well-thought structure for a program of
research and data collection rather than for a single experiment. The
data are no longer unidimensional, and the research goals are
multifaceted. Data sets are multidimensional and may be compiled from
many sources. The questions on this larger scale to investigate a
physical science or engineering problem can be where in the overall plan
to gather/use data, where to employ statistical principles, where to
simulate, when and how to verify. In the social sciences it may be a
matter of understanding how to combine public and proprietary sources of
data, how to link time sequences of events and data, how to define
multi-person decision-making (independent, adversarial, cooperative,
informed/uninformed). Once again, \textit{before data}. Obviously, with
data the more familiar work is underway.

\section*{Challenge \#5---To Use Available Data to Advance Education in
Statistics}

\subsection*{Statistics Education $\cap$ Embedded Statistical~Software}

These new large shared databases are being used as primary research
resources and as teaching tools for undergraduate and graduate
\textit{science courses}. By providing some statistical methodology that
is integrated into the database resource, the database creators provide
the scientific impetus as well as a prime opportunity for researchers
and students alike to explore the richness of [multidisciplinary]
information and to discover interesting relationships within the
databases. Faculty report that use of the western North American
volcanic and intrusive rocks NAVDAT database (Walker et al., \citeyear{Walker2006}) in
undergraduate geology to analyze students' own conjectures has
popularized research as a curricular activity and has created an
unforeseen enthusiasm among geology students for data analysis with the
relatively simple internal statistical methodology. (It is not at all
clear that this taste of statistics has also created a hunger for
high-dimensional or other more advanced statistical methods, although
the high-dimensional data suggests their great potential.) The scope of
SkyServer (Sloan Digital Sky Survey/SkyServer,\break available at \url{http://skyserver.sdss.org})
is even broader with an
open invitation to the public to explore and to analyze astronomy data.

At present these explorations are limited primarily by the
sophistication of the readily available (internal) statistical software
and the researcher's intuition or the student's inquisitiveness.
Incorporation of more extensive statistical tools, both for data
exploration and for modeling, could educate the science students in the
power of sound statistical analytic methods, both simple and advanced.
Simultaneously, utilizing these large scientific databases in statistics
classes allows \textit{primary} investigation of interdisciplinary
questions and application of\break exploratory, high-dimensional and/or other
advanced statistical methods by going beyond textbook data sets.

One challenge to the statistical community is to identify opportunities
and mechanisms to incorporate statistical software that is equally as
sophisticated as the scientific information in these large resource
databases.

A different challenge to the statistical community is to take advantage
of these rich scientific data sources and the opportunities they provide
for individual investigations in statistics courses. Textbooks' focus on
end-of-the-chapter problems and on oft-used data sets have served to
assist students to mimic research investigations, with prespecified
questions and data that is well-adapted to analysis by a specific
methodology. These public sources of complex data open new possibilities
to make the statistical investigation of conjectures exciting, even at
rather basic levels. Both NAVDAT and SkySurvey suggest directions to
explore the information from the level of elementary school to
post-doctoral research. The challenge for statistical education will be
to do this as well.

Still the most stunning statement by scientists and researchers during
the workshop was the query: What  does Statistics have to do with data
availability? And why would Statisticians care---apart from the
policies of their own professional journals?

The answer is implicit in these Challenges that emerged from this highly
multidisciplinary group of very thoughtful individuals as they expressed
the difficulties and the successes in making data publicly available and
usable in each of their disciplines. The purpose of this paper is to
initiate a discussion of these important issues within the statistical
community. In addition, these issues need to be examined in each
scientific discipline and ultimately find their way into the training of
scientists.

\section*{Acknowledgments}

We thank the following speakers at the workshop for bringing insights to
the policies and practices faced by journals in striving for openness in
scientific research and research reporting. By framing the issues, both
philosophical and practical, they enabled a thoughtful, vigorous
dialogue:

Rolf Apweiler, European Bioinformatics Institute of the European
Molecular Biology Laboratory;

Richard W. Carlson, Co-Editor, \textit{Earth and Planetary Science
Letters}, Carnegie Institution of Washington;

Katrina L. Kelner, Deputy Editor for Life Sciences, \textit{Science},
American Association for the Advancement of Science;

Gary King, Principal Investigator, \textit{The Dataverse Network
Project}, Harvard University;\

Robert A. Moffitt, Chief Editor, \textit{American Economic Review},
Johns Hopkins University;

Jordan Raddick, \textit{Sloan Digital Sky Survey}, Johns Hopkins
University;

Ani Thakar, \textit{Sloan Digital Sky Survey}, Johns Hopkins
University;

Edward J. Ungvarsky, Capital Defender, Northern Virginia Capital
Defender Office, Arlington, Virginia.

\end{document}